\documentclass[showpacs,preprintnumbers,prd,amsmath,amssymb]{revtex4}
\usepackage{amsmath}
\usepackage{graphicx}
\usepackage{subfig}
\usepackage{hyperref}
\usepackage{amssymb}
\usepackage[english]{babel}
\usepackage{epsfig}
\usepackage{wasysym}
\usepackage{comment}
\newcommand{\be}{\begin{equation}}
	\newcommand{\ee}{\end{equation}}
\newcommand{\bea}{\setlength\arraycolsep{2pt} \begin{eqnarray}}
	\newcommand{\eea}{\end{eqnarray}}

\def\0{{\sst{(0)}}}
\def\1{{\sst{(1)}}}
\def\2{{\sst{(2)}}}
\def\3{{\sst{(3)}}}
\def\4{{\sst{(4)}}}
\def\5{{\sst{(5)}}}
\def\6{{\sst{(6)}}}
\def\7{{\sst{(7)}}}
\def\8{{\sst{(8)}}}
\def\sst#1{{\scriptscriptstyle #1}}

\begin{document}

\title{Black holes in $f(R)$ gravity coupled to nonlinear electromagnetic fields}

\author{Yong-Ben Shi \footnote{Corresponding author}}
\affiliation{College of Physics Science and Technology, Hebei University, Baoding 071002, China}

\author{Rong-Jia Yang \footnote{Corresponding author}}
\email{yangrongjia@tsinghua.org.cn}
\affiliation{College of Physics Science and Technology, Hebei University, Baoding 071002, China}
\affiliation{Hebei Key Lab of Optic-Electronic Information and Materials, Hebei University, Baoding 071002, China}
\affiliation{National-Local Joint Engineering Laboratory of New Energy Photoelectric Devices, Hebei University, Baoding 071002, China}
\affiliation{Key Laboratory of High-pricision Computation and Application of Quantum Field Theory of Hebei Province, Hebei University, Baoding 071002, China}

\begin{abstract}
We establish a framework to construct spherically symmetric and static solutions in $f(R)$ gravity coupled with nonlinear electromagnetic fields. We present two new specific solutions and discuss the energy conditions. We calculate some thermodynamic quantities of the obtained black holes like entropy and energy and investigate their thermodynamic topology.
\end{abstract}
\keywords{BH, $f(R)$ gravity, Nonlinear electrodynamics}

\maketitle

\section{INTRODUCTION}
\label{sec1}
General Relativity (GR), as a preeminent theory of gravitation, has substantiated by a plethora of experimental and observational validations. Recent evidences \cite{SNIa,SNIa1,SNIa2,CMB,CMB2,BAO,BAO1,BAO2,BAO3,BAO4,LSS,LSS1} have unveiled a perplexing revelation: the universe is currently undergoing a phase of accelerated expansion. This cosmic acceleration challenges the conventional understanding of gravity, necessitating the presence of an exotic form of matter that violates the strong energy condition (SEC) within the framework of GR. This enigmatic component, commonly referred to as dark energy, has prompted extensive discussions within the scientific community. The most renowned contender for explaining dark energy is the cosmological constant, often interpreted as a manifestation of quantum vacuum energy. However, a substantial discrepancy between its observed and predicted values persists, spanning approximately 120 orders of magnitude. This discrepancy underscores the need for alternative frameworks to elucidate the nature of dark energy.

Modified gravity models provide alternative explanations for dark energy, such as $f(R)$ gravity \cite{Sotiriou:2008rp,Nojiri:2010wj,Capozziello:2011et,Johnson:2019vwi,Li:2007xn,Gogoi:2020ypn,DeFelice:2010aj,Kruglov_2023,Kruglov:2018dir,Makarenko:2014lxa,Kruglov:2015aqm,Kruglov:2014rja}, $f(R,T)$ theory \cite{fRT,fRT1,fRT2}, $f(\mathcal{G})$ gravity \cite{Nojiri:2005jg,Li:2007jm}, $f(R,\mathcal{G})$ gravity \cite{fRG,fRG2,fRG3,fRG5}, $f(R,T_{\mu\nu}T^{\mu\nu})$ theory \cite{Katirci:2013okf}, and quantum fluctuation modified gravity \cite{Dzhunushaliev:2013nea,Yang:2015jla,Liu:2016qfx}, where $\mathcal{G}$ represents the Gauss-Bonnet term, $T_{\mu\nu}$ and $T$ respectively denotes the energy-momentum tensor and its trace, and $R_{\mu\nu}$ is the Ricci tensor. Among various modified gravity models, $f(R)$ gravity has garnered widespread attentions in the field of high-curvature gravity theories. It involves replacing the Ricci scalar $R$ in the Einstein-Hilbert action with an arbitrary function of $R$  

GR and $f(R)$ gravity have some common solutions, but there are also solutions in $f(R)$ gravity that differ from those in GR. These distinct solutions possess different physical properties, highlighting the significance of finding black hole (BH) solutions in $f(R)$ gravity. However, due to the presence of high-order derivative terms in the equations of motion within $f(R)$ gravity, solving these equations becomes challenging. Despite the complexity, many solutions in $f(R)$ gravity have been obtained, see for example \cite{Multamaki:2006zb,Capozziello:2007id,Capozziello:2012iea,AnhKy:2018pbb,DeFalco:2023twb,VanKy:2022itq,Ky:2024lce,delaCruz-Dombriz:2009pzc,Hendi:2011hxq,Nashed:2021sey,Nashed:2021mpz,Nashed:2021lzq,Nashed:2021ffk,Nashed:2020mnp,Nashed:2020kdb,Nashed:2020tbp,Tang:2019qiy,Nashed:2018oaf,Nashed:2018efg,Nashed:2018piz,Amirabi:2015aya, Moon:2011hq}. These solutions serve as a testing ground for the predictions and implications of $f(R)$ gravity, allowing researchers to evaluate its compatibility with observational data and theoretical consistency. Despite the challenges posed by the complexity of the equations of motion, the pursuit of exact BH solutions in $f(R)$ gravity remains a significant area of research, offering the potential to deepen our understanding of gravity and its modifications at cosmological and astrophysical scales. 

Nonlinear electrodynamics was firstly proposed in \cite{Born:1934gh}, which removes a singularity of point-like charges. There are many models
of nonlinear electrodynamics, such as Born–Infeld-like \cite{Kruglov_2010,Kruglov:2016uzf}, logarithmic \cite{Soleng:1995kn}, 
exponential electrodynamics \cite{Hendi:2012zz,Kruglov:2016lqd}, Heisenberg-Euler-type \cite{Kruglov:2007bh} and others. Many BH solutions in GR coupled with nonlinear electromagnetic fields have been found, see for example \cite{Kruglov:2019okd,Ayon-Beato:1998hmi,Kruglov:2017ymn,Ayon-Beato:1999kuh,Kruglov:2023cyb,Kruglov:2017mpj,Fan:2016hvf}. BH solutions in $f(R)$ gravity coupled to nonlinear electromagnetic fields have also been investigated widely, see for example \cite{Nashed:2024lem,Hurtado:2020gic,Nashed:2022xmv,Rodrigues:2015ayd,Olmo:2011zz,Rodrigues:2016fym,Nojiri:2017kex,HabibMazharimousavi:2011yj,Olmo:2011ja,Guerrero:2020uhn,EslamPanah:2024tex,Sekhmani:2024gft}. In this paper, we will construct new solutions of spherically symmetric and static BH within the framework of $f(R)$ gravity coupled to nonlinear electromagnetic fields and investigate the physical properties of these solutions.

The paper is organized as follows. In Sec. \ref{s2}, we will introduce the fundamental equations for $f(R)$ theory, which is minimally coupled to nonlinear electromagnetic fields, adopting a spherically symmetric static metric. In Sec. \ref{s3}, we will present a framework for solving the equations of motion in $f(R)$ gravity and give two specific solutions of BH. In Sec. \ref{s4}, we will consider the energy conditions (ECs) in $f(R)$ gravity and provide their specific expressions for the obtained solutions. In Sec. \ref{s5}, we will discuss the thermodynamic topology and provide the winding number and topological charge for the two BH solutions.  In Sec. \ref{s6}, we make final remarks and draw conclusions.

%%%%%%%%%%%%%
\section{The equations of motion in $f(R)$ gravity}
\label{s2}
%%%%%%%%%%%%%%%%%

We consider $f(R)$ gravity coupled to nonlinear electromagnetic fields as follows
\begin{eqnarray}
I=\frac{1}{16\pi} \int d^4 x \sqrt{-g}[f(R)-\mathcal{L}(F)]
\label{action}, 
\end{eqnarray}
where $g$ is the determinant of the metric, $f(R)$ is an analytic function of $R$, $ F=F_{\mu\nu} F^{\mu\nu}$ with $F_{\mu\nu}=\partial_\mu A_\nu-\partial_\nu A_\mu$ being the Faraday-Maxwell tensor, and the Lagrangian
density $\mathcal{L}( F)$ is a function of $ F$. Throughout the paper, We take the units $G=c=1$.

Applying the variational principle in terms of the metric $g_{\mu\nu}$ and the potential $A_\mu$ to the action \eqref{action}, the resulting equations are respectively given by
\begin{equation}
 f_{R}  R_{\mu \nu } - \frac{1}{2} g_{ \mu \nu } f+\left ( g_{\mu \nu} \square -  \triangledown_\mu \triangledown_\nu \right) f_{R}  = 
 T_{\mu \nu } \ , \\ \label{31}
\end{equation}
\begin{equation}
\triangledown_\mu\left(\mathcal{L}_F F^{\mu \nu}\right) = 0 \ ,\\ \label{32}
\end{equation}
where $\mathcal{L}_F =\frac{\partial\mathcal{L}}{\partial F }$, $ f_{R}=\frac{df(R)}{dR}$ and the energy-momentum tensor is 
$T_{\mu \nu}= 2\left(\mathcal{L}_F  F^2_{\mu\nu}- \frac{1}{4} g_{\mu\nu}\mathcal{L}\right)$.

Considering a spherically symmetric and static space-time, whose element line in Schwarzschild coordinates reads
\begin{equation}
    ds^2= -e^{m(r)} dt^2 + e^{n(r)} dr^2 + r^2(d\theta^2 + sin\theta^2
 d\phi^2) \ , \label{metric}
\end{equation}
where $m(r)$ and $n(r)$ are analytic functions of the radial coordinate $r$, and assuming that
\begin{equation}
    A = a(r) dt + Q_m cos\theta d \phi \ , \label{vf}
\end{equation}
where $Q_m $ is the total magnetic charge defined by
$Q_m = \frac{1}{4\pi} \int F $, we obtain the Ricci scalar as 
\bea
R=\frac{e^{-n} [-2 r^2 m''+r m' (r n'-4)-r^2 m'^2+4 r n'+4 e^{n}-4]}{2 r^2} \label{rs} \ ,
\eea
where $n\equiv n(r)$, $m\equiv m(r)$, $m'=\frac{dm(r)}{dr}$, $m''=\frac{d^2m(r)}{d^2r}$ and $n'=\frac{dn(r)}{dr}$. Using metric \eqref{metric}, Ricci scalar \eqref{rs}, and vector field \eqref{vf} with an assumption $a(r)=0$, we find that Eqs. \eqref{31} and \eqref{32} can be written as 
\bea
&&\frac{1}{2} e^{m} (f- \mathcal{L})+\frac{f_{R} e^{m-n} \left[2 r m''+m' (4-r n')+r m'^2\right]}{4 r}-e^{m-n} f_{R}''+\frac{1}{2} e^{m-n} f_{R}' \left(  n'-\frac{4 }{r}\right)=0 \label{m1} \ ,\\
&&\frac{1}{2} e^{n} \left( \mathcal{L} - f \right)+\frac{(r m'+4) f_{R}'}{2 r}+\frac{f_{R} (-2 r m''+r m' n'-r m'^2+4 n')}{4 r} = 0\ , \label{m2}\\
&&\frac{1}{2} \left(r^2 \mathcal{L}-\frac{4 Q_m^2 \mathcal{L}_ F }{r^2}-r^2 f\right)+\frac{1}{2} r e^{-n} f_{R}' (r m'-r n'+2)+\frac{1}{2} f_{R} e^{-n} (-r m'+r n'+2 e^{n}-2)+r^2 e^{-n} f_{R}'' = 0 \ , \label{m3}
\eea
where $f_{R}'=\frac{df_{R}}{dr}$ and $f_{R}''=\frac{d^2f_{R}}{d^2r}$. Thus for metric \eqref{metric}, we obtain the equations of motion for $f(R)$ gravity coupled to non-linear electrodynamics, denoted as Eqs. \eqref{m1} , \eqref{m2}, and \eqref{m3}, respectively.

%%%%%%%%%%%%%%%%%%%%
\section{New Black hole Solutions}
\label{s3}
%%%%%%%%%%%%%%%%%%%%%%%%%

In this section, we will use Eqs. \eqref{m1}-\eqref{m3} to derive spherically symmetric and static solutions of BH in the $f(R)$ gravity coupled to nonlinear electrodynamics fields. The
exploration of spherically symmetric and static BH solutions can provide valuable insights into the behavior of gravity
within the framework of $f(R)$ theories.

Dividing Eq. \eqref{m1} with $e^{-n(r)+m(r)}$, and adding or subtracting \eqref{m2}, we obtain respectively
\bea
&&\frac{(r f_{R}'+2 f_{R}) (m'+n')}{2 r}-f_{R}''=0 \ , \label{mm1} \\
&&-e^{n} (\mathcal{L}-f)+f_{R} m''-\frac{1}{2} f_{R} m' n'+\frac{1}{2} f_{R} m'^2+\frac{f_{R} m'}{r}-\frac{f_{R} n'}{r}+\frac{1}{2} f_{R}' n' -\frac{1}{2} f_{R}' m' -\frac{4 f_{R}'}{r}-f_{R}''=0 \ . \label{mm2}
\eea
From Eq. \eqref{mm1}, we have
\bea
n'(r)=-m'(r)+\frac{2 r f_{R}''(r)}{r f_{R}'(r)+2 f_{R}(r)}\ . \label{mpp}
\eea
We observe that if the functions $f_{R}$ and $m'$ in Eq. \ref{mpp} are given, we can calculate $n'$. On the other hand, if $m'$ and $n'$ are given, we can compute the function $f$ in terms of $r$ by using Eq. \eqref{rs} as follows
\bea
f(r)= \int f_{R}(r) dR = \int f_{R}(r) \frac{dR}{dr} dr \ . \label{f1}
\eea

 From Eq. \eqref{mm2}, yields
\bea
  \mathcal{L}= \frac{1}{2 r}e^{-n} \left(2 r f_{R} m''-r f_{R} m' n'+r f_{R} m'^2+2 f_{R} m'+2 r f e^{n}+r f_{R}' n'-8 f_{R}'-2 f_{R} n'-r f_{R}' m' -2 r f_{R}''\right)\ . \label{LL1}
\eea 
Inserting Eq. \eqref{LL1} into Eq. \eqref{m3}, we can determine that 
\bea
\mathcal{L}_F =\frac{r^2 e^{-n}}{8 Q_m^2} \left[f_{R} \left(r^2 (2 m''-m' n'+m'^2)+4 (e^{n}-1)\right) + r f_{R}' (r m'-r n'-4)+2 r^2 f_{R}''\right] \ . \label{LLFF1} 
\eea
Note that there is no term containing $f$ in Eq. \eqref{LLFF1}, we can use Eqs. \eqref{LL1} and \eqref{LLFF1} to verify the solutions we obtained.

The square of the field strength \(  F  \) is given by
\bea
 F =\frac{2 Q_m^2}{r^4} .
\eea
Therefore, we can express \( \mathcal{L} \) and \( \mathcal{L}_ F  \) as functions of the field strength \(  F \). 

Next, we discuss how to choose an appropriate \( f_{R} \) such that Eq. \eqref{mpp} can be integrated. Because the quantities that needs to be solved exceed the number of equations, we need additional assumptions about $f(R)$. According to the spherically symmetry of the metric, we consider such a type of $f(R)$ which fulfills \(2 f_{R}(r) = c_2 r f'_{R}(r) \) with \( c_2 \) a nonzero constant. Substituting it into Eq. \eqref{mpp}, we can obtain
\bea
n'(r)= - m'(r) + \frac{2 f_{R}''(r)}{(c_2+1) f_{R}'(r)} \ ,
\eea
in which $c_2\neq -1$. Integrating both sides, we get 
\bea
n(r)= - m(r) + \frac{2 \ln \left[m_0f_{R}'(r)\right]}{c_2+1} + \mathcal{I}\ . \label{nn}
\eea
where \(\mathcal{I}\) is the constant of integration, we set \(\mathcal{I} = 0\) and $m_0=1$ here for simplicity without losing generality. 

Solving the differential equation, \(  2 f_{R}(r) = c_2 r f'_{R}(r)\), we find
\bea
f_{R}(r) = c_1 \left(\frac{r}{m_1}\right)^{\frac{2}{c_2}} = c_1 \left(\frac{r}{m_1}\right)^{c_3} \ , \label{fRnn}
\eea
where \( c_1 \) is a constant and $c_3 = 2/c_2$, \( m_1 \) is an integral constant with a dimension of mass, which can be written as a linear combination of mass and magnetic charge. Since GR is a special case of $f(R)$ theory, namely, the spherically symmetric and static metric \eqref{metric} contains Schwarzschild BH as a special solution, we can assume that \( e^{m(r)} = 1 - \frac{2 M(r)}{r} \) and find by using Eqs. \eqref{nn} and \eqref{fRnn}
\bea
e^{n(r)}&=& \frac{r \left(\frac{r}{m_1}\right){}^{\frac{2 \left(c_3-1\right) c_3}{c_3+2}}}{r-2 M(r)} \ .
\eea
Then the Ricci scalar can be expressed as
\bea
R&=&\frac{2 \left(\frac{r}{m_1}\right){}^{-\frac{2 \left(c_3-1\right) c_3}{c_3+2}}}{\left(c_3+2\right) r^3} \bigg[\left(c_3+2\right) r \left(\frac{r}{m_1}\right){}^{\frac{2 \left(c_3-1\right) c_3}{c_3+2}}+c_3 r^2 M''(r)-c_3^2 r M'(r)+3 c_3 r M'(r)\cr
&&\quad-3 c_3^2 M(r)+3 c_3 M(r)+2 c_3^2 r-3 c_3 r+2 r^2 M''(r)+4 r M'(r)-2 r\bigg]. \label{Rn} 
\eea
Rewriting Eq. \eqref{f1}  as 
\bea
f(r) &=& \frac{c_1}{\left(c_3+2\right){}^2} \int \frac{1}{r^4}\left(\frac{r}{m_1}\right){}^{-\frac{\left(c_3-4\right) c_3}{c_3+2}} \bigg[-4 c_3^2 r \left(\frac{r}{m_1}\right){}^{\frac{2 \left(c_3-1\right) c_3}{c_3+2}}-16 r \left(\frac{r}{m_1}\right){}^{\frac{2 \left(c_3-1\right) c_3}{c_3+2}}-16 c_3 r \left(\frac{r}{m_1}\right){}^{\frac{2 \left(c_3-1\right) c_3}{c_3+2}}-8 c_3^4 r\cr
&&\quad+12 c_3^3 r-8 c_3^2 r+24 c_3 r+16 r+\left(2 c_3^2 r^3+8 c_3 r^3+8 r^3\right) M^{(3)}(r)+\left(-6 c_3^3 r^2-4 c_3^2 r^2+20 c_3 r^2+8 r^2\right) M''(r)\cr
&&\quad+\left(4 c_3^4 r-18 c_3^3 r-14 c_3^2 r-12 c_3 r-32 r\right) M'(r)+\left(12 c_3^4-6 c_3^3+30 c_3^2-36 c_3\right) M(r)\bigg] \, dr    \label{Frn} \ ,
\eea
and utilizing Eqs. \eqref{LL1} and \eqref{LLFF1}, we derive 
\bea
\mathcal{L}(r)&=& f(r)+\frac{2 c_1 \left(\frac{r}{m_1}\right){}^{-\frac{\left(c_3-4\right) c_3}{c_3+2}}}{\left(c_3+2\right) r^3} \left[c_3 \left(2 c_3+1\right) \left(r \left(M'(r)-2\right)+3 M(r)\right)-\left(c_3+2\right) r^2 M''(r)\right] \ , \label{Lrn}\\
\mathcal{L}_ F (r)&=&\frac{c_1 r \left(\frac{r}{m_1}\right){}^{-\frac{\left(c_3-4\right) c_3}{c_3+2}} }{2 \left(c_3+2\right) Q_m^2}
\bigg[\left(c_3+2\right) r \left(\frac{r}{m_1}\right){}^{\frac{2 \left(c_3-1\right) c_3}{c_3+2}}-4 c_3 r-2 r+\left(-c_3 r^2-2 r^2\right) M''(r)\cr
&&\quad+\left(4 r-c_3 r\right) M'(r)+9 c_3 M(r)\bigg]\label{lf} \ .
\eea
Analyzing these above equations, we find that if Eq. \eqref{Frn} can be integrated, all the required physical quantities can be expressed in analytical forms. So, we now analyze what is the form of \( M(r) \) that allows Eq. \eqref{Frn} to be integrated. To be more insightful, we rewrite Eq. \eqref{Frn} as
\bea
f(r)&=&\int dr \mathcal{K} \left(\mathcal{K}_1+\mathcal{K}_2 \right)\ ,
\eea
where
\bea
\mathcal{K}&=&\frac{c_1}{\left(c_3+2\right){}^2}  \frac{1}{r^4}\left(\frac{r}{m_1}\right){}^{-\frac{\left(c_3-4\right) c_3}{c_3+2}}
 \ ,\\
\mathcal{K}_1&=&-4 c_3^2 r \left(\frac{r}{m_1}\right){}^{\frac{2 \left(c_3-1\right) c_3}{c_3+2}}-16 r \left(\frac{r}{m_1}\right){}^{\frac{2 \left(c_3-1\right) c_3}{c_3+2}}-16 c_3 r \left(\frac{r}{m_1}\right){}^{\frac{2 \left(c_3-1\right) c_3}{c_3+2}}-8 c_3^4 r\cr
&&\quad+12 c_3^3 r-8 c_3^2 r+24 c_3 r+16 r
\ , \\
\mathcal{K}_2&=&\left(2 c_3^2 r^3+8 c_3 r^3+8 r^3\right) M^{(3)}(r)+\left(-6 c_3^3 r^2-4 c_3^2 r^2+20 c_3 r^2+8 r^2\right) M''(r)\cr
&&\quad+\left(4 c_3^4 r-18 c_3^3 r-14 c_3^2 r-12 c_3 r-32 r\right) M'(r)+\left(12 c_3^4-6 c_3^3+30 c_3^2-36 c_3\right) M(r) \ ,
\eea
where $\mathcal{K}$ includes all terms that do not contain \( M \) or its derivatives, \(\mathcal{K}_2\) includes the product of the power of \( r \) and the derivative of \( M(r) \) with respect to \( r \), where $M^{(3)}(r)=d^3M/dr^3$ and \( M(r) \) can be seen as the zero order derivative of  \( M(r) \). \(\mathcal{K}_1\) includes the remaining terms. 

Due to 
\bea
\int \mathcal{K} \left(\frac{r}{m_1}\right)^{c_{4}} dr= -\frac{c_1\left(\frac{r}{m_1}\right){}^{c_3 \left(\frac{6}{c_3+2}-1\right)+c_{4}}}{\left(c_3+2\right) \left[c_3^2-c_3-\left(c_3+2\right) c_{4}+6\right] r^3},
\eea
were \( c_{4} \) is a constant which is not equal to $c_3+\frac{12}{c_3+2}-3$. For \( c_{4}=c_3+\frac{12}{c_3+2}-3 \), we have
\bea
\int \mathcal{K}  \left(\frac{r}{m_1}\right)^{c_{4}} dr&=& \frac{c_1\ln \left(\frac{r}{m_1}\right)}{\left(c_3+2\right){}^2 m_1^3} \ .
\eea
Therefore, \( \int dr \mathcal{K} \left(\frac{r}{m_1}\right)^{c_{4}} \) can be integrated for any constant \( c_{4} \). 

%%%%%%%%%%%%%%%%%%%%
\subsection{Solution A}
\label{n2}
%%%%%%%%%%%%%%%%%%%%%%%%%

Now we consider a simple case by setting \( c_1 = 1 \) and \( c_3 = 3 \) in Eq. \eqref{fRnn}, i.e., \( f_{R} = \left(\frac{r}{m_1}\right)^3 \), and assume that \( f(R) = m_1^{-2} \ln( p m_1^2R) \),  where \( p \) is an nonzero constant and satisfies $pR>0$. In section V, we will determine the value of \( p \) by investigating the thermodynamics topology of BH. In asymptotically safe gravity, an effective action containing $\ln(R)$ has been obtained \cite{Bonanno:2012jy}. It is appropriate to consider such a term here. Using Eqs. \eqref{Rn} and \eqref{Frn}, we obtain
\bea
&&\frac{2}{25 r^4} \left(\frac{r}{m_1}\right){}^{3/5} \left[r \left(-25 r^2 M^{(3)}(r)+65 r M''(r)+178 M'(r)+154\right)-486 M(r)\right] +\frac{4}{m_1^3}\cr
&&\quad+\frac{m_1^2 \left[r \left(25 r^2 M^{(3)}(r)-65 r M''(r)-178 M'(r)-154\right)+486 M(r)\right]-50 r^3 \left(\frac{r}{m_1}\right){}^{2/5}}{5 m_1^2 \left[m_1^2 r \left(r \left(5 r M''(r)+4 M'(r)+7\right)-18 M(r)\right)+5 r^4 \left(\frac{r}{m_1}\right){}^{2/5}\right]}=0.
\eea
Solving the above differential equation, yields
\bea
M(r)= \frac{25 \left(13 m_1-6 r\right) \left(\frac{r}{m_1}\right){}^{12/5}}{1092}+\frac{r}{2}+\frac{\mathcal{D}_1}{r^{9/5}}+\mathcal{D}_2 r^2 \label{Mn1},
\eea
or
\bea
M(r)=\frac{r}{2}-\frac{25 r}{182}  \left(\frac{r}{m_1}\right){}^{12/5}+\mathcal{D}_1 r^{27/5}+\frac{\mathcal{D}_2}{r^{9/5}}+\mathcal{D}_3 r^2\label{Mn2},
\eea
were \( \mathcal{D}_1 \), \( \mathcal{D}_2 \), and \( \mathcal{D}_3 \) are constants. For solution \eqref{Mn2}, the Ricci scalar \( R \) equals a constant. However, for the case where \( R \) is a constant, the steps we established above will fail, namely solution \eqref{Mn2} is unpysical. Solution \eqref{Mn1} is somewhat complicated, it is difficult for us to obtain further analytical results from it. So we additionally take \( \mathcal{D}_1 = 0 \) and \( \mathcal{D}_2 = 0 \) in Eq. \eqref{Mn1}, deriving

\bea
e^{m(r)}&=&\frac{25 \left(6 r-13 m_1\right) \left(\frac{r}{m_1}\right){}^{12/5}}{546 r}\ ,\label{sl11} \\
e^{n(r)}&=&\frac{546 r}{25 \left(6 r-13 m_1\right)} \ ,\label{sl12}\\
R&=&\frac{m_1}{r^3} \ ,\label{r1} \\
f(R)&=&\frac{ \ln{\left(\frac{m_1^3p}{r^3}\right)} }{m_1^2}=\frac{\ln{(pm_1^2 R)}}{m_1^2} \ , \label{fr1}\\
\mathcal{L}( F )&=& \frac{7 m_1 \left[2 \ln \left(\frac{m_1^3 p  F ^{3/4}}{2^{3/4} \left(Q_m^2\right){}^{3/4}}\right)+23\right]-72  \left(\frac{2 Q_m^2}{ F }\right){}^{1/4}}{14 m_1^3} \ ,\label{FrL3}\\
\mathcal{L}_ F ( F )&=&\frac{3 \left[12  \left(\frac{ 2 Q_m^2}{ F }\right)^{1/4}+7 m_1\right]}{28  F  m_1^3} \ . \label{FrLF30}
\eea
Using Eqs. \eqref{Mn1} and \eqref{lf}, we obtain the result for \( \mathcal{L}_ F  \), which matches Eq. \eqref{FrLF30}, implying that our results are correct.

\subsection{Solution B}
\label{n3}
Now we consider the second simple case where \( f(R) = m_1^{-2} \ln(p m_1^2R) \) and \( f_{R} = \left(\frac{r}{m_1}\right)^{-3} \) by setting \( c_1 = 1 \) and \( c_3 = -3 \) in Eq. \eqref{fRnn}. Using Eqs. \eqref{Rn} and \eqref{Frn}, we find
\bea
&&\left[-2 m_1^{24}+r^{24} \left(r^2 M^{(3)}(r)+37 r M''(r)+344 M'(r)-550\right)+756 r^{23} M(r)\right] \times\cr\quad 
&&\frac{\Big[-m_1^{19} r^5+2 m_1^{24}+2 r^{24} \left(r M''(r)+14 M'(r)-25\right)+72 r^{23} M(r)\Big]}{m_1 r \left[m_1^{24}+r^{24} \left(r M''(r)+14 M'(r)-25\right)+36 r^{23} M(r)\right]}=0\ .\label{de2}
\eea
Solving this equation, we get
\bea
M(r) = \frac{m_1^{19} \left(19 r^5-18 m_1^5\right)}{4788 r^{23}}+\frac{\mathcal{D}_4}{r^9}+\frac{\mathcal{D}_5}{r^4}+\frac{r}{2} \label{M-3},
\eea
or
\bea
M(r) = -\frac{m_1^{24}}{266 r^{23}}+\frac{\mathcal{D}_4}{r^{21}}+\frac{\mathcal{D}_5}{r^9}+\frac{\mathcal{D}_6}{r^4}+\frac{r}{2},\label{m21}
\eea
where \( \mathcal{D}_4 \), \( \mathcal{D}_5 \), and \( \mathcal{D}_6 \) are integration constants. For the same reasons as in the previous subsection, solution \eqref{m21} is unpysical. We additionally take \( \mathcal{D}_4 = \frac{1}{2}m_1^{10} \) and \( \mathcal{D}_5 = 0 \) in Eq. \eqref{M-3} to get analytic functions, obtaining
\bea
e^{m(r)}&=&-\frac{m_1^{10} \left(19 m_1^9 r^5-18 m_1^{14}+2394 r^{14}\right)}{2394 r^{24}} \ ,\label{sl21} \\
e^{n(r)}&=&-\frac{2394 m_1^{14}}{19 m_1^9 r^5-18 m_1^{14}+2394 r^{14}} \ ,\label{sl22}\\
R&=&\frac{r^3}{m_1^5} \ , \label{r2}\\
f(R)&=&\frac{\ln\left({\frac{pr^3}{m^3_1}}\right)}{m_1^2}=\frac{ \ln{(p m_1^2R)}}{m_1^2} \ ,\label{fr12} \\
\mathcal{L}( F )&=& \frac{266 \ln \left[\frac{2^{3/4} p \left(Q_m^2\right){}^{3/4}}{m_1^3  F ^{3/4}}\right]+\frac{126\ 2^{3/4} m_1^5  F ^{5/4}}{\left(Q_m^2\right){}^{5/4}}-247}{266 m_1^2}
\ ,\label{FrL-3}\\
\mathcal{L}_ F ( F )&=&\frac{45 ( F /2)^{1/4} m_1^3 }{38  \left(Q_m^2\right){}^{4/5}}-\frac{3}{4  F  m_1^2}
\ . \label{FrLF3}
\eea
Using Eqs. \eqref{M-3} and \eqref{lf}, we get \( \mathcal{L}_ F  \) which matches Eq. \eqref{FrLF3}, also implying that our results are correct.

\section{Energy conditions}
\label{s4}
To ensure that a given solution satisfies certain physical conditions, we consider the so-called ECs. In order to appropriately express these conditions, we reformulate Eq. \eqref{31} in terms of an effective energy-momentum tensor \( T_{\mu\nu}^{(\rm{eff})} \) as follows.
\bea
R_{\mu \nu} - \frac{1}{2} R g_{\mu \nu } = \frac{1}{f_{R}}\left[T_{\mu \nu } + \frac{1}{2} g_{\mu \nu}(f - f_{R} R) - (g_{\mu \nu}\square -  \triangledown_\mu \triangledown_\nu)f_{R}\right]\equiv T_{\mu\nu}^{\rm{(eff)}}.\label{tef}
\eea
The term \( T_{\mu\nu}^{\rm{(eff)}} \) as defined above signifies the effective energy-momentum tensor arising from $f(R)$ gravity, functioning as the source term in Einstein's equations. This tensor comprises the canonical energy-momentum tensor of the matter fields, \( T_{\mu\nu} \), scaled by \( f^{-1}_{R} \), in addition to contributions from the nonlinear $f(R)$ function in the Lagrangian density. These modifications to the source terms in Einstein's equations suggest potential changes in the ECs when compared to those in GR.
 
Making the identifications 
\bea
\label{encd}
&&T^{\rm{(eff)}0}_0 = - \rho^{\rm{(eff)}},\\
\label{encd1}
&&T^{\rm{(eff)}1}_1 = p_{\rm{r}}^{\rm{(eff)}}, \\
\label{encd2}
&&T^{\rm{(eff)}2}_2 =  T^{\rm{(eff)}3}_3 = p_{\rm{t}}^{\rm{(eff)}}, 
\eea
the ECs can be expressed as follows (for ECs in GR, see for example \cite{visser}; for studies related to ECs in \( f(R) \) theories, see for example \cite{santos,santos1,santos3} )
\bea
&&{\rm{NEC_{1,2}}}:~~ \rho^{\rm{(eff)}} + p_{\rm{{r,t}}}^{\rm{(eff)}} \geq 0 \ , \label{NEC}\\
&&{\rm{SEC}}:~~ \rho^{\rm{(eff)}} + p_{\rm r}^{\rm{(eff)}} + 2p_{\rm t}^{\rm{(eff)}} \geq 0 \ ,\\
&&{\rm{WEC_{1,2}}}:~~ \rho ^{\rm{(eff)}} + {p_{\rm{{r,t}}}}^{\rm{(eff)}} \geq 0 \ ,\\
&&{\rm{WEC_3=DEC_{1}}}:~~\rho^{\rm{(eff)}} \geq 0 \ , \\
&&{\rm{DEC_{2,3}}}:~~ \rho^{\rm{(eff)}} - p_{\rm{{r,t}}}^{\rm{(eff)}} \geq 0 \label{DEC}\ , 
\eea
where NEC, WEC, DEC denotes the null energy condition, the weak energy condition, and the dominant energy condition, respectively. 

Now we check whether the solutions obtained above satisfy the ECs in \( f(R) \) gravity. We first consider solution A. Substituting Eqs. \eqref{sl11}, \eqref{sl12}, \eqref{r1}, \eqref{fr1},  \eqref{encd}, \eqref{encd1}, and \eqref{encd2} into the ECs \eqref{NEC}–\eqref{DEC}, we find that
\bea
&&{\rm{NEC_1}=\rm{WEC_1}}:~~\frac{10 \left(6 r-13 m_1\right)}{91 r^3} \ , \\
&&{\rm{NEC_2}= \rm{WEC_2}}:~~\frac{3 \left(13 m_1+68 r\right)}{182 r^3} \ ,\\
&&{\rm{SEC}}:~~\frac{132 r-91 m_1}{91 r^3} \ , \\
&&{\rm{DEC_1}=\rm{WEC}_3}:~~\frac{66}{91 r^2} \ , \\
&&{\rm{DEC_2}}:~~\frac{2 \left(65 m_1+36 r\right)}{91 r^3} \ , \\
&&{\rm{DEC_3}}:~~\frac{60 r-39 m_1}{182 r^3} \ .
\eea
All ECs are satisfied if $ r\geq  \frac{13m_1}{6} $, which is imposed by the $ \rm{WEC}_1  \geq 0 $. Since $ r_{+} = \frac{13m_1}{6} $, where $r_+$ denotes the radius of the event horizon, so the energy conditions are not satisfied behind the event horizon.

For solution B, inserting  Eqs. \eqref{sl21}, \eqref{sl22}, \eqref{r2}, \eqref{fr12},  \eqref{encd}, \eqref{encd1}, and \eqref{encd2} into the energy conditions \eqref{NEC}–\eqref{DEC}, yields the energy conditions
\bea
&&{\rm{NEC_1}}={\rm{WEC_1}}:~~\frac{24 r^{12}}{m_1^{14}}+\frac{4 r^3}{21m_1^5}-\frac{24}{133 r^2} \ , \\
&&{\rm{NEC_2}}={\rm{WEC_2}}:~~\frac{18 r^{12}}{m_1^{14}}-\frac{r^3}{2 m_1^5}+\frac{276}{133 r^2} \ ,\\
&&{\rm{SEC}}:~~\frac{30 r^{12}}{m_1^{14}}-\frac{19 r^3}{21 m_1^5}+\frac{264}{133 r^2} \ , \\
&&{\rm{DEC_1}}={\rm{WEC_3}}:~~\frac{19 r^5 \left(m_1^9+315 r^9\right)+396m_1^{14}}{399m_1^{14} r^2} \ , \\
&&{\rm{DEC_2}}:~~\frac{6 r^{12}}{m_1^{14}}-\frac{2 r^3}{21 m_1^5}+\frac{288}{133 r^2} \ , \\
&&{\rm{DEC_3}}:~~\frac{12 r^{12}}{m_1^{14}}+\frac{25 r^3}{42 m_1^5}-\frac{12}{133 r^2} \ .
\eea
The NEC$_{2}$, WEC$_{2,3}$, SEC, and DEC$_{1,2}$ conditions hold true throughout the entire spacetime. NEC$_1$ and WEC$_1$ are fulfilled for \( r \geq 0.69573 m_1 \), while DEC holds for \( r \geq 0.6430 m_1 \). In other words, all energy conditions are valid outside the event horizon ($r_+\simeq 0.69573m_1$, see the next section).

\section{THERMODYNAMICS TOPOLOGY OF THE black hole }
\label{s5}
In this section, we will explore the thermodynamic topological properties of the BH obtained above. We first introduce the method to calculate thermodynamic quantities in a spherically symmetric and static space-time. For the Hawking temperature \( T \), we have \cite{Hendi:2010gq,Cognola:2011nj}
\bea
T= \frac{\kappa}{2 \pi} =\frac{\sqrt{\mathcal{M}^{'}(r_+)\mathcal{N}^{'}(r_+)}}{4\pi}, \, \label{TT}
\eea
where \(\kappa\) represents the surface gravity, \(\mathcal{M} = e^{m(r)}\), and \(\mathcal{N} = e^{-n(r)}\). The entropy and the energy of BH are respectively given by \cite{Zheng:2019mvn,Zhu:2020hte}
\bea
S &=& \pi r_+^2 f_{R} \ , \label{SS} \\
E &=& \frac{1}{2} \int^{r_+}\sqrt{\frac{\mathcal{M}^{'}}{\mathcal{N}^{'}}}\bigg[ \frac{f_{R}}{f_{R}^{'}}+\frac{1}{2} (f - R f_{R} ) \bigg] r^2 dr \ . \label{EE}
\eea
The generalized free energy of a BH with arbitrary mass is defined as \cite{Wei:2022dzw}
\bea
\mathcal{F}=E-\frac{S}{\tau} \ , \label{FF}
\eea
where the parameter \( \tau \) is an additional variable with the dimension of time that can vary freely. It can be considered as the inverse of the temperature of the cavity surrounding the BH. In general, this generalized free energy is off-shell, except when \( \tau = 1/T \).

Using the generalized free energy, we can construct a vector field \cite{Wei:2022dzw}
\bea
\phi=(\phi^r,\phi^\Theta)=\bigg( \frac{\partial\mathcal{F}}{\partial r_+},-\cot{\Theta} \csc{\Theta} \bigg) \ ,
\eea
where \( \Theta = \frac{\pi}{2} \) and \( \tau = 1/T \) are the zero points of this vector field. The topological characteristics linked to the zero points of the field are defined by its winding number \( w \) or the topological charge \( W = \sum_i w_i \), where $w_i$ is the winding number for the $i$-th zero point of the vector field.

To calculate the winding number, we begin by constructing a contour \( C \) around each zero point, parameterized as \cite{Hazarika:2024cpg}
\bea
\left\{
\begin{aligned}
r_+ &= a \cos{\mathcal{B}}  +r_0\\
\Theta &= a_1 \sin{\mathcal{B}} + \frac{\pi}{2} \ ,
\end{aligned}
\right. \label{c1f}
\eea
where the parameter \( \mathcal{B} \) ranges from \( 0 \) to \( 2\pi \). By calculating the deflection \( \Omega \) of the vector field \( n \) along the contour \( C \), we find that the winding number is given by 
\bea
w = \frac{1}{2 \pi} \Omega(2 \pi) =\frac{1}{2 \pi} \int_0^{2\pi} \epsilon_{12} n^1 \partial_{\mathcal{B}} n^2 d\mathcal{B} \ ,
\eea
where
\bea
n^1=\frac{\phi^r}{\sqrt{(\phi^r)^2+(\phi^\Theta)^2}} \ \ \ {\rm and} \ \ \ n^2=\frac{\phi^\theta}{\sqrt{(\phi^r)^2+(\phi^\Theta)^2}} \ .
\eea
This method for computing the topological number or charge is referred to as Duan’s $\phi$ mapping technique \cite{Duan:1984ws,Duan:1979ucg}.

\subsection{Thermodynamics topology of the BH solution  A }

For solution A, we find that the largest positive roots of \( \mathcal{M} \) and \( \mathcal{N} \) are the same. Solving \( 
\mathcal{M}= 0 \), we have
\bea
m_1 = \frac{6 r_+}{13} \ , \label{jie}
\eea
Using Eqs. \eqref{TT}, \eqref{SS}, \eqref{EE}, \eqref{FF}, and \eqref{jie}, we can calculate the temperature $T$ and \( \frac{\partial\mathcal{F}}{\partial r_+} \) as 
\bea
T &=& \frac{25 \sqrt[5]{\frac{13}{6}}}{168 \pi  r_+} \label{Tr} \ ,\\ 
\frac{\partial\mathcal{F}}{\partial r_+} &=&\frac{2197 \sqrt[5]{\frac{13}{6}} \left[3 \ln \left(\frac{216 p}{2197}\right)+10\right]}{2592}-\frac{2197 \pi  r_+}{108 \tau } \ .\label{FFr+}
\eea
Inserting these two equations into \( \frac{\partial\mathcal{F}}{\partial r_+} =0 \) with \( \tau = \frac{1}{T} \), yields
\bea
p = \frac{2197}{216 e^{15/7}} \ .
\eea
Thus, the value of \( p \) in Eqs. \eqref{FrL3} and \( f(R) = m_1^{-2} \ln(p m_1^2R) \) is now determined.

The diagram for the components \( \phi_r \) and \( \phi_\theta \) of  vector field $\phi$ denoted by arrows are shown in Figure 1 (a), with \( \tau = \frac{168}{25} \sqrt[5]{\frac{6}{13}} \pi  r_0\). We observe that the zero point of the vector field is located at \( (r_+/r_0 = 1, \Theta = \pi/2)\). In Figure 1 (b), we plot \(r_+\) as a function of \( \tau \), where $r_+$ increases monotonically with $\tau$. 
\begin{figure}[h]
    \centering
    \begin{minipage}[b]{0.30\textwidth}
        \centering
        \includegraphics[width=\textwidth]{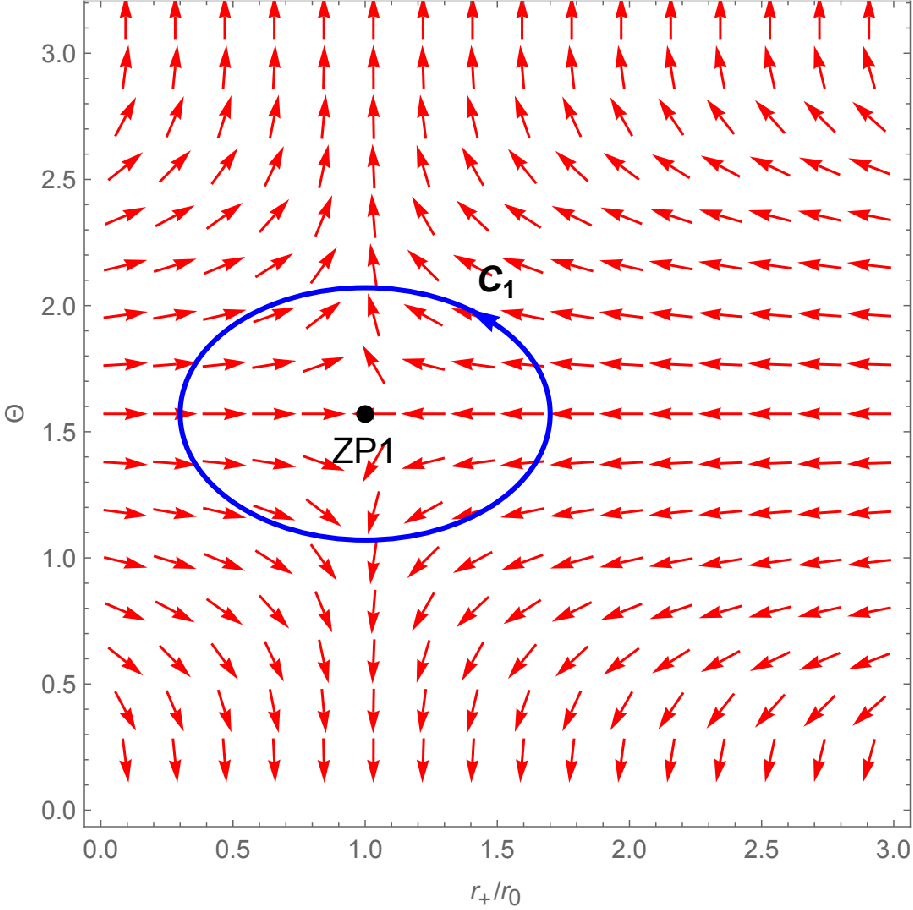}
        \caption*{(a)}
        \label{1}
    \end{minipage}
     \hspace{3em} 
    \begin{minipage}[b]{0.40\textwidth}
        \centering
        \includegraphics[width=\textwidth]{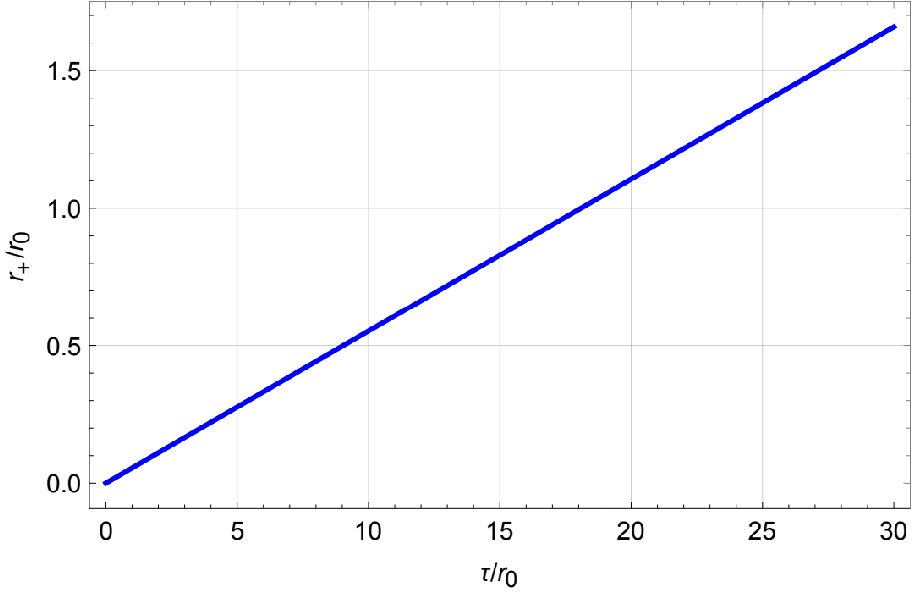}
        \caption*{(b)}
        \label{2}
    \end{minipage}
\caption{In figure (a), \(\Theta\) as a function of \(r_+\). In figure (b),  \(r_+\) as a function of \(\tau\).}
\end{figure}
Since the winding number \( w \) is independent of the loops enclosing the zero point, we can calculate it by using any loop, such as \( C_1 \) with \( a = 0.7 r_0 \) and \( a_1 = 0.5 \) in Eqs. \eqref{c1f}, as shown in figure 1 (a). After some calculations, we find that the winding number is \( w = -1 \). Because there is only one zero point, the topological charge \( W \) is also \( -1 \).

\subsection{thermodynamics topology of the BH solution  B }

For solution B, \( \mathcal{M} \) and \( \mathcal{N} \) also share the same roots, but we are unable to derive any analytical solutions for  \( \mathcal{M}=\mathcal{N}=0 \). After observing and analyzing the figure 2 where we plot \(\mathcal{N}\) as a function of \(r/m_1\), we numerically get the solution as \(r_+\simeq 0.69573m_1 \).
% 1.43734
\begin{figure}[h]
    \centering
    \begin{minipage}[b]{0.45\textwidth}
        \centering
        \includegraphics[width=\textwidth]{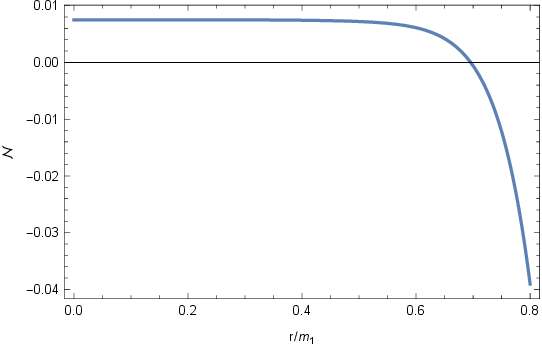}
       
        \label{3}
    \end{minipage}
    \caption{$\mathcal{N}$ as a function of $r/m_1$. }
\end{figure}

Using the same method as in the previous section, we obtain
\bea
T &=&\frac{0.579281}{r} \ ,\\
\frac{\partial\mathcal{F}}{\partial r_+}&=&9.40911 \ln (p)-\frac{18.6579 r_+}{\tau }+95.7972 \ , \\
p &=& 0.000119448 \ . 
\eea
Therefore, we determine the value of \( p \) in Eqs. \eqref{FrL-3} and \( f(R) = m_1^{-2} \ln(p m_1^2R) \) .

We show the components \( \phi_r \) and \( \phi_\theta \) of vector field \(\phi\) denoted by the arrows with \( \tau = 1.72628 r_+\) in figure 3 (a). The zero point locates at \( (r_+/r_0 = 1, \Theta = \pi/2)\). \(r_+\) as a function of \( \tau \) is plotted in figure 3 (b), where $r_+$ also increases monotonically with $\tau$.

Utilizing loop \(C_2\), whose parametric equation is the same of loop \(C_1\), we also find that both the winding number and the topological charge are \(-1\).

\begin{figure}[h]
    \centering
    \begin{minipage}[b]{0.30\textwidth}
        \centering
        \includegraphics[width=\textwidth]{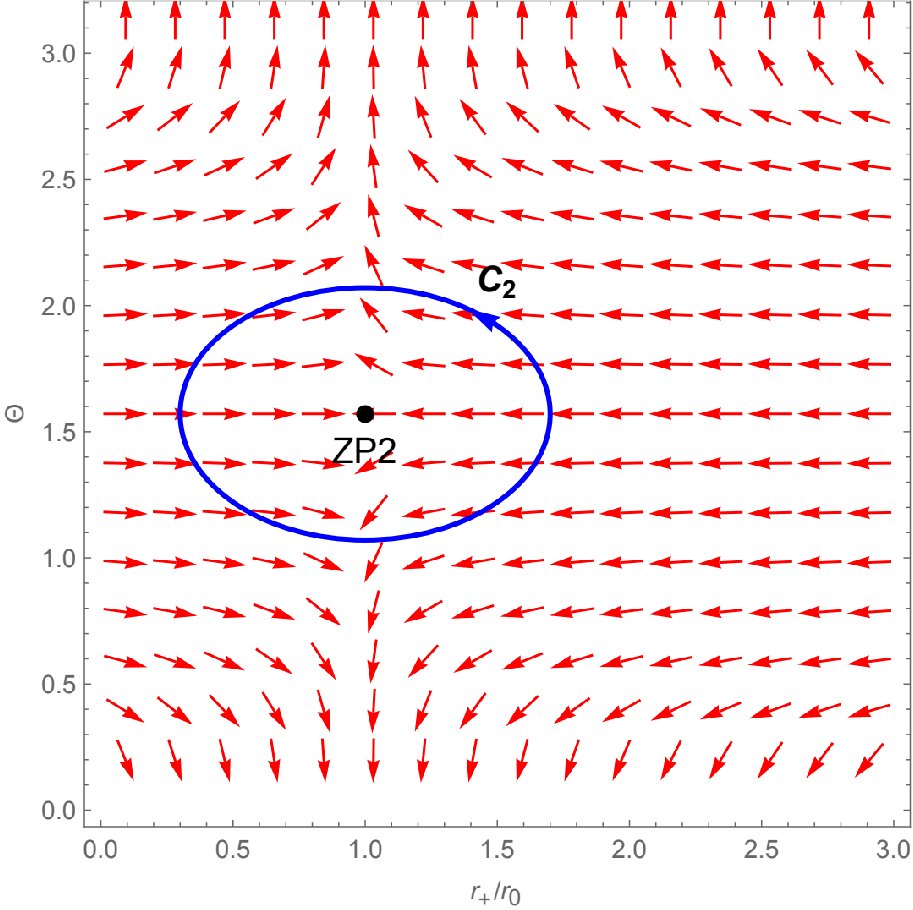}
        \caption*{(a)}
        \label{4}
    \end{minipage}
     \hspace{3em} % 使用 \hfill 在两个 minipage 之间添加可伸缩的空白
    \begin{minipage}[b]{0.40\textwidth}
        \centering
        \includegraphics[width=\textwidth]{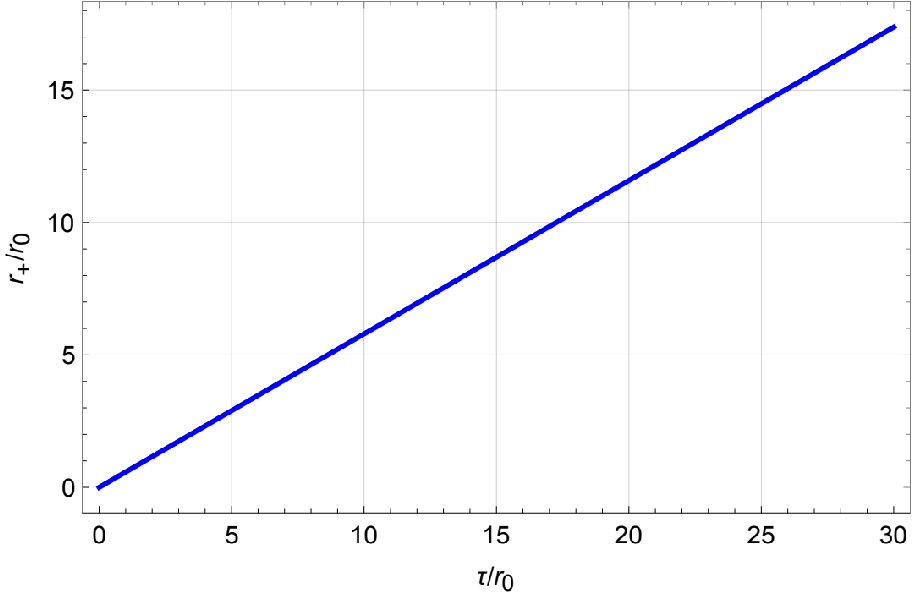}
        \caption*{(b)}
        \label{5}
    \end{minipage}
    \caption{In figure (a), \(\Theta\) as a function of \(r_+\). In figure (b),  \(r_+\) as a function of \(\tau\).}
\end{figure}

\section{CONCLUSION and discussions}
\label{s6}
We discussed solutions in \( f(R) \) gravity coupled to nonlinear electromagnetic sources. We introduced a framework for solving \( f(R) \) gravity: first, we assume the form of \( f_{R}(r) \) with respect to \( r \) and solve \( f(r) \) in the form of an integral; next assuming an analytic form of \( f(R) \) with respect to \( R \) and substituting the expression of \( R \) into it, then one can obtain the solutions by using \( \frac{df(R)}{dr} = \frac{df(r)}{dr} \).

As applications, we explicitly presented two BH solutions: one with assumptions \( f_{R} = \left(\frac{r}{m_1}\right)^3 \) and \( f(R) = m_1^{-2}\ln(pm_1^2 R) \), the other with assumptions \( f_{R} = \left(\frac{r}{m_1}\right)^{-3} \) and \( f(R) = m_1^{-2}\ln(pm_1^2 R) \). We analyzed the energy conditions and found that if \( r \geq r_+ \) all energy conditions could hold for both BH solutions. We also discussed the thermodynamic topological properties of these two solutions and found that the winding number and the topological charge for both solutions are equal to \(-1\).

As a possible development for future research, we can also solve equation \eqref{mm1} for other forms of \( f_{R} \). For example, taking \( m' + n' = -\frac{1}{b_0 r} \) with \( b_0 \) an constant and substituting it into equation \eqref{mm1}, yields
\[
2 b_0 r^2 f_{, R}''(r)+r f_{R}'(r)+2 f_{R}(r)=0 \ .
\]
Solving this differential equation provides a new form of \( f_{R} \). Additionally, we can consider other expressions for \( f(R) \), such as \( f(R) = m_1^{-2} \ln(p - p m_1^2 R) \). Of course, we can also directly focus on \( f(r) \) by assuming specific forms for \( e^{m(r)} \) and \( f_{R}(r) \), then solving for \( R(r) \) to obtain \( r(R) \) inversely, substituting it back into \( f(r) \) , which will also give us \( f(R) \).

\begin{acknowledgments}
This study is supported in part by National Natural Science Foundation of China (Grant No. 12333008) and Hebei Provincial Natural Science Foundation of China (Grant No. A2021201034).
\end{acknowledgments}

\textbf{Data Availability Statement}: No Data associated in the manuscript.

\bibliographystyle{apsrev}
\bibliography{name}

\end{document}